\documentclass[aps,prd,superscriptaddress,nofootinbib,preprintnumbers]{revtex4-1}
\usepackage{amsmath}
\usepackage{graphicx}
\usepackage{subfigure}
\usepackage{amssymb}
\usepackage{xcolor}
\usepackage{multirow}
\usepackage{cancel}
\usepackage{color}
\usepackage{ulem}
\usepackage{caption}
\usepackage{listings}
\usepackage{xcolor}
\usepackage{float}
\usepackage{slashed}

\usepackage{graphicx}

\usepackage{float}
\usepackage{amsmath}
\usepackage{amssymb}
\usepackage[utf8]{inputenc}
\usepackage{hyperref}
\usepackage{color}
\usepackage{slashed}
\usepackage[absolute]{textpos}
\usepackage{changes}
\allowdisplaybreaks

\graphicspath{{images/}}

\newcommand{\be}{\begin{equation}}
\newcommand{\ee}{\end{equation}}

\begin{document}

\begin{textblock}{4}(0.2,0.1)   
  \includegraphics[width=2cm]{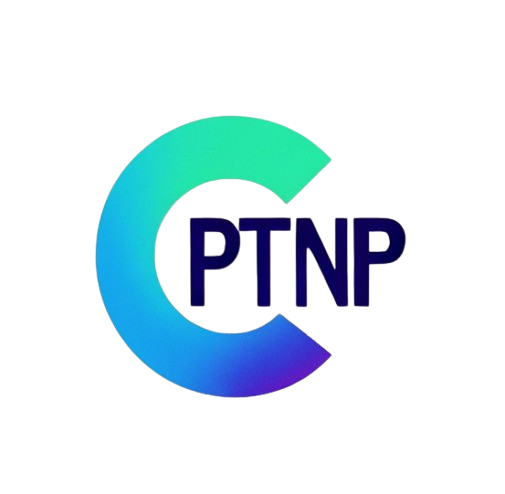}  
\end{textblock}

\begin{textblock}{6}(9,0.5)
  \raggedleft
  {\text{CPTNP-2025-021}}  
\end{textblock}

\title{Jet Reconstruction with Mamba Networks in Collider Events}

\author{Jinmian Li}
\email{jmli@scu.edu.cn}
\affiliation{College of Physics, Sichuan University, Chengdu 610065, China}

\author{Peng Li}
\email{lipeng@scu.edu.cn}
\affiliation{College of Physics, Sichuan University, Chengdu 610065, China}

\author{Bingwei Long}
\email{bingwei@scu.edu.cn}
\affiliation{College of Physics, Sichuan University, Chengdu 610065, China}
\affiliation{Southern Center for Nuclear-Science Theory (SCNT), Institute of Modern Physics, Chinese Academy of Sciences, Huizhou 516000, Guangdong Province, China}

\author{Rao Zhang}
\email{rzhang9527@gmail.com}
\affiliation{College of Physics, Sichuan University, Chengdu 610065, China}

\begin{abstract}
We introduce a novel end-to-end framework for jet reconstruction in high-energy collider events, leveraging the efficiency and long-range modeling capabilities of the Mamba architecture.
Our model unifies instance segmentation, classification, and kinematic regression into a single multi-task learning system, enabling a sophisticated multi-level reconstruction that simultaneously identifies primary heavy jets ($t$, $H$, $W/Z$) and their constituent sub-jets. 
To facilitate supervised learning for this complex task, we develop a novel method for assigning final-state hadrons to their ancestor colored partons using a Mixed-Integer Linear Programming solver, which generates high-fidelity ground-truth labels. 
The model achieves high classification accuracy, with an Average Precision score of 0.569 for $W/Z$-jets and 0.568 for $b$-jets, and shows exceptional precision in kinematic reconstruction. 
Furthermore, we show that the model not only maintains stable performance in high-pileup environments but also successfully reconstructs the mass peaks of beyond the standard model particles. 
This work presents a powerful and versatile new tool for comprehensive event reconstruction at the LHC.
\end{abstract}


\maketitle

\newpage
\section{Introduction}\label{sec:intro} 
In high-energy collisions at facilities like the Large Hadron Collider (LHC), the fundamental interactions produce quarks and gluons. However, the principle of color confinement prevents these partons from being directly observed. Instead, they rapidly evolve into streams of stable, detectable hadrons through the complex processes of parton showering and hadronization. These emergent, collimated structures, known as jets, serve as the primary experimental signatures of the underlying quarks and gluons. The central challenge of jet physics is therefore to reverse this process: to accurately reconstruct jets from the myriad of final-state particles and reliably infer the kinematic properties of the initial partons. A successful reconstruction algorithm must be able to map the observed hadronic energy flow back to the four-momentum of its progenitor parton in a way that is insensitive to the soft and collinear details of the shower and hadronization, ensuring a stable bridge between theoretical prediction and experimental measurement.

At the LHC, jets are conventionally reconstructed using sequential recombination algorithms~\cite{Salam:2010nqg}, with the anti-$k_T$ algorithm~\cite{Cacciari:2008gp} serving as the standard workhorse for both the ATLAS and CMS collaborations. The choice of a radius parameter, $R$, is critical: a smaller cone (e.g., $R=0.4$) is effective for jets from light quarks and gluons, while the hadronic decays of boosted heavy particles like top quarks, $W/Z$ boson, or Higgs bosons produce ``fat jets" that require a much larger cone size (with $R \sim 2m/p_T$) to contain their decay products. These fat jets exhibit characteristic internal substructures, which led to the development of a suite of ``jet substructure" techniques~\cite{Abdesselam:2010pt,Altheimer:2012mn,Altheimer:2013yza,Adams:2015hiv,Larkoski:2017jix,Kogler:2018hem} designed to identify them. Powerful examples include mass-drop tagging~\cite{Butterworth:2008iy} for the Higgs boson, HEPTopTagger algorithm~\cite{Plehn:2010st} for the top quark, and N-subjettiness~\cite{Thaler:2010tr,Thaler:2011gf} for general fat-jets.
A significant practical challenge, however, particularly for large-$R$ jets, is the heavy contamination from pileup. To combat this, grooming techniques such as trimming~\cite{Krohn:2009th}, pruning~\cite{Ellis:2009su}, and soft drop declustering~\cite{Larkoski:2014wba} were developed to remove soft, wide-angle radiation. More advanced methods like pileup per particle identification~\cite{Bertolini:2014bba} use particle-level information to probabilistically remove pileup contributions. Despite these advances, precisely reconstructing the parent parton's momentum remains a challenge, and the performance of these algorithms is a critical component of precision physics programs at the LHC.

In recent years, Machine Learning (ML) has emerged as a revolutionary paradigm, with deep learning-based taggers substantially outperforming traditional substructure techniques in classification tasks, as detailed in several community reviews~\cite{Larkoski:2017jix,Guest:2018yhq,Albertsson:2018maf,Radovic:2018dip,Larkoski:2024uoc}. The rich information within a jet can be represented in various forms amenable to different neural network architectures~\cite{deOliveira:2015xxd,Komiske:2016rsd,Kasieczka:2017nvn,Macaluso:2018tck,Moreno:2019bmu,Komiske:2018cqr,Qu:2019gqs,Louppe:2017ipp,Cheng:2017rdo,Andreassen:2018apy,Semlani:2023kzf,Ai:2024mkl,Hammad:2024cae,Mikuni:2025tar,Jahin:2025plg}. Early pioneering work treated jets as images, applying Convolutional Neural Networks (CNNs) to calorimeter grids~\cite{deOliveira:2015xxd,Komiske:2016rsd}. More recently, the field has gravitated towards more natural, irregular representations like point clouds of constituent particles, processed by Graph Neural Networks (GNNs), with models like ParticleNet~\cite{Qu:2019gqs} setting new performance benchmarks. The state-of-the-art has further advanced with the application of attention-based mechanisms, leading to models like Particle Transformer (ParT)~\cite{pmlr-v162-qu22b}. Concurrently, a new class of ``theory-informed" networks has emerged that explicitly build in fundamental physical symmetries, such as Lorentz group equivariance~\cite{Brehmer:2024yqw}. Models like LorentzNet\cite{Gong:2022lye} and PELICAN~\cite{Bogatskiy:2022czk} achieve exceptional performance by designing network layers that respect the underlying physics. 
These models achieve typical Area Under the Curve (AUC) values exceeding 0.98 for top tagging, without considering pileup effects. Furthermore, the momentum reconstruction component of the PELICAN network predicts the $p_T$ and mass of the $W$ boson with a standard deviation of a few percent.

Despite their remarkable success, a common limitation of these taggers is their reliance on a traditional clustering algorithm for the initial jet-finding step. Consequently, they inherit any distortions or inefficiencies from the initial jet definition. To overcome this, a promising new direction treats jet reconstruction as an end-to-end computer vision task, learning to identify jets directly from all particles in an event. 
In Monte-Carlo simulations of collider events, final-state hadrons can be unambiguously traced back to their color-neutral ancestor parton. This principle allows for the application of supervised learning to train neural networks for identifying jet constituents from final-state particles. For instance, Ref.~\cite{Ju:2020tbo} utilizes a supervised GNN to construct $W$ boson jets. In our work~\cite{Guo:2020vvt}, we enhanced the GNN with a focal loss function, ensuring its efficiency is maintained even in the presence of significant pileup. Furthermore, we demonstrated that this GNN, when trained on $H$+jets events, can generalize to detect Higgs jets in various other processes. An alternative approach, image segmentation with CNNs, has also proven effective for identifying Higgs and top quark jets simultaneously in event images. Following this paradigm, in Ref.~\cite{Li:2020grn,Choi:2023slq} we represent event data as digital images and apply the Mask R-CNN framework~\cite{2017arXiv170306870H} to reconstruct the Higgs and top quark jet. Collectively, these deep learning approaches yield higher efficiencies in jet detection and greater accuracy in momentum reconstruction compared to traditional jet clustering and substructure tagging algorithms.
However, these approaches face their own challenges. CNNs can struggle to capture long-range correlations across an image—a feature that is critical for identifying decay products that are spatially separated or jets split across the periodic boundaries of the detector image. Furthermore, extending these models to simultaneously detect multiple, potentially overlapping, and distinct jet types in complex final states presents a significant hurdle.

In this work, we propose a novel framework, SAsmJM (Segment Any sm-Jet Model), that moves beyond both traditional CNNs and computationally intensive Transformers by leveraging the power of Mamba~\cite{gu2023mamba}, a recent and highly efficient architecture based on state space models. While Vision Transformers~\cite{dosovitskiy2020image} have shown success in applying global attention to entire images, their quadratic complexity with respect to the number of image patches can be a bottleneck. Mamba offers an alternative path, capturing long-range dependencies in data with linear complexity, 
making it exceptionally well suited for a global, event-level view of jet reconstruction. 
We treat the entire event as a single image in the pseudorapidity-azimuth ($\eta - \phi$) plane and employ a vision-adapted Mamba network, 
VMamba-V2~\cite{liu2024vmamba}, as the core feature extractor. 
This approach allows the model to effectively establish connections between distant energy deposits—such as those from a Higgs boson decaying to spatially separated b-jets or from a jet split across the periodic $\phi$
boundary—problems that pose a challenge for the limited receptive fields of standard CNNs. Our model is designed as a multi-task learning system, performing simultaneous instance segmentation, classification, and kinematic regression for five distinct jet categories: top quark, Higgs boson, $W/Z$ boson, bottom quark, and light-quark/gluon jets. Crucially, this framework enables a multi-level reconstruction, allowing the model to not only identify a parent particle like a top quark but also to recognize its subsequent decay products ($W$ and $b$-jet inside the top jet). This hierarchical approach, unlike traditional taggers that provide a single label for a fat jet, offers a more granular and physically complete picture of the event. We train the network on a diverse set of Standard Model (SM) processes and demonstrate its robustness against significant pileup contamination. Finally, to validate that the model has learned generalizable physical principles rather than process-specific features, we evaluate its performance on unseen Beyond the Standard Model (BSM) events, showcasing its potential as a powerful and versatile tool for event reconstruction at the LHC. 

The remainder of this paper is organized as follows. Section~\ref{sec2} details our Monte-Carlo event simulation and data pre-processing techniques. Section~\ref{sec3} presents the neural network architecture and its key hyperparameters. In Section~\ref{sec4}, we assess the model's baseline performance on classification and momentum regression tasks. To demonstrate the robustness and generality of our method, Section~\ref{sec5} evaluates its performance under varying pileup conditions and on several BSM signal processes. We summarize our findings and conclude in Section~\ref{sec6}.

\section{Data preparation} \label{sec2}
\subsection{Event simulation}

Event samples for the training and testing of our neural network are generated using the \texttt{MG5\_aMC@NLO} framework~\cite{madgraph}. Subsequent parton showering, hadronization, and particle decays were handled by \texttt{Pythia8}~\cite{pythia}. No detector simulation is performed. To approximate the angular granularity of the calorimeters, final-state stable particles are projected onto a fixed-resolution $\eta-\phi$ grid, as detailed later.

We simulate a variety of the SM processes initiated by proton-proton collisions at a center-of-mass energy of $\sqrt{s}=13$ TeV. The simulated final states include:

\begin{align}
jjjj, b\bar{b}WW, Hb\bar{b}, HHWW, Ht\bar{t}, tb, t\bar{t}, t\bar{t}W, t\bar{t}Z, t\bar{t}t\bar{t}, WH, WW, WZ, ZH, ZZ, ZZZZ~.
\end{align}
Since the primary objective of this work is to identify boosted $H$, $t$, $W$, and $Z$ bosons that form fat jets, the transverse momentum ($p_T$) cuts of $p_T^{b/W/Z/H} > 200 \, \text{GeV}$ and $p_T^t > 300 \, \text{GeV}$ were applied at parton level during the event generation. Furthermore, these heavy particles were required to decay hadronically: $t \to Wb$, $H \to b\bar{b}$, $W/Z \to jj$. 
For each process, a sample of 10000 events was generated for training. For the final performance evaluation, independent test and validation datasets were created. Each set contains 1000 events for each of the 16 processes, generated with the same simulation pipeline but a different random seed to ensure statistical independence. 

A realistic simulation must account for pileup events, the multiple proton-proton collisions that occur in each bunch crossing at the LHC. These pileup interactions are predominantly low-momentum-transfer, non-diffractive events. Their simulation relies on phenomenological models to describe complex processes, including parton showering, Lund-string hadronization, multiple-parton interactions, and color reconnection. The parameters of these models are not uniquely determined and must be tuned to match the experimental data. A specific, validated set of these parameters is known as a ``tune"~\cite{ATLAS:2012uec}. For this work, we simulate pileup events using the A3 tune of Pythia8, whose parameters are detailed in Refs.~\cite{Skands:2014pea,ATLAS:2016puo}. The number of pileup interactions per bunch crossing is Poisson-distributed.  For our study, we assume a constant mean of $\langle \mu \rangle=50$ pileup interactions. 

\subsection{Assignment of final-state hadrons to ancestor quarks}
Due to color confinement during hadronization, it is generally difficult to unambiguously trace a final-state hadron back to its parent quark, a difficulty that persists even in Monte Carlo simulations. This challenge is rooted in the dynamics of hadronization, as described by models like the Lund String Model. In these frameworks, hadrons form from the fragmentation of color strings stretched between partons, allowing a single hadron in a complex event to inherit momentum from multiple string segments. This process complicates a direct, one-to-one mapping from hadron to parton. While a perfect solution to this ambiguity is challenging, our work provides a method to recover the kinematic information of the initial-state quarks with high accuracy by identifying a single, globally optimal assignment for all ambiguous particles.
We classify final-state hadrons into two categories based on their ancestral traceability. Hadrons that can be unambiguously mapped to a single parent quark constitute the unique state; their set is denoted by \textbf{U}$_k$, with $k$ indices for the ancestor quark. Conversely, hadrons whose momentum is collectively inherited from a group of color-connected ancestor quarks form the common state; their set is denoted by \textbf{C}.
The momentum of the $k$-th initial-state quark is then reconstructed using the following assignment rule:
\begin{equation}
p_k = \sum_{u \in \textbf{U}_k} p_u + \sum_{c \in \textbf{C}} \alpha_{kc} p_c,
\end{equation}
where $k \in \{1, 2, \dots, K\}$ indexes the initial-state quarks. The first term aggregates the momenta of all hadrons $p_u$ from the corresponding unique-state set $\textbf{U}_k$. The second term distributes the momenta of all common-state hadrons $p_c$ from the set \textbf{C}. This distribution is governed by the binary assignment coefficients $\alpha_{kc}$, which are defined as 
\begin{equation}
  \alpha_{kc} =
  \begin{cases}
    1, & \text{if common-state hadron $c$ is assigned to quark $k$} \\
    0, & \text{otherwise}
  \end{cases}~.
\end{equation}

Our approach to determining the assignment matrix $\alpha_{kc}$ is motivated by two intertwined considerations. From a practical standpoint, a hard assignment ($\alpha_{kc} \in \{0, 1\}$) is essential to define the unambiguous ground-truth labels needed for effective segmentation training. From a physical standpoint, this simplification is guided by the Lund String Model. The model suggests that it is improbable for momentum contributions from competing string configurations to be perfectly balanced, implying that a ``dominant" source parton often exists.

The existence of a dominant source provides the physical justification for seeking a binary assignment matrix. However, our method does not simply enforce a greedy, local assignment of each hadron to its pre-identified dominant source, as this may not yield the optimal reconstruction of the ancestor quarks' momenta globally. Instead, our method is holistic: its central goal is to find the single, complete assignment matrix $\alpha_{kc}$ that allows the reconstructed quark momentum, $p_k$, to best approximate the true momentum, $p_{q_k}$, across the entire event. To find the optimal assignment, we seek to minimize a loss function defined as the total absolute difference between the true and reconstructed momenta:
\begin{align}
\mathcal{L}(\alpha)= \sum_{k=1}^{k=K} \Delta p_k \equiv \sum_{k=1}^{k=K} |p_{q_k} - p_k (\alpha_{kc})|~.~
\end{align}
This minimization is framed as a constrained optimization problem. Specifically, we aim to find the assignment matrix $\alpha$ that solves
\begin{equation}
\begin{aligned}
\hat{\alpha} = \arg\min_{\alpha} ~\mathcal{L}(\alpha),\quad 
\text{subject to} ~\sum_{k=1}^{K} \alpha_{kj} = 1, ~ \forall j \in \mathbf{C}, ~ \alpha_{kj} \in {0, 1}.
\end{aligned}
\end{equation}
The constraint ensures that each common-state hadron $j$ is assigned to exactly one ancestor quark. This formulation can be interpreted as finding the single hadronization assignment that best satisfies the global momentum conservation of the event. It corresponds to a Mixed-Integer Linear Programming (MILP) problem, which we solve using a dedicated solver. We define a successful reconstruction for quark $k$ as one where the individual error $\Delta p_k$ is below a threshold of 5 GeV. These successful assignments are stored as ground-truth labels for training our network. 
A key advantage of this method is its generalizability: the assignment logic is not limited to initial-state partons but can be applied to any intermediate parent particle within a decay chain. Consider, for instance, the process $pp \to Ht\bar{t} \to (b\bar{b})(bW^+)(\bar{b}W^-) \to (b\bar{b})(bjj)(\bar{b}jj)$. By targeting different stages of this cascade, a single simulated event can furnish ground-truth labels for a multitude of jet types, including the $H$-jet (from $H\to b\bar{b}$), two top-jets (from $t\to bW$), $W$-jets (from $W\to j j$) and the subsequent four $b$-jets and four light-flavor jets. This strategy creates a diverse and densely labeled dataset from a single process, which is crucial for training a robust model capable of simultaneously identifying various jet signatures.

The MILP assignment process is a simulation-level tool that requires access to the full Monte Carlo truth tree. In experimental data, individual hadrons cannot be resolved from calorimeter deposits, making the inverse reconstruction problem ill-posed. Accordingly, this procedure is designed exclusively for generating ground-truth labels and is not intended for direct application to experimental data.
This limitation, however, is confined to the training data preparation stage. Once trained, the neural network learns a direct mapping from detector-like inputs (in our case, particles projected onto a finite-resolution grid) to the ideal, MILP-derived labels. During inference, the network therefore operates solely on these energy deposits, entirely independent of the MC truth information and the MILP solver.

\subsection{The inputs to the neural network}
We represent each event as a multi-channel image by projecting its final-state particles onto a grid in the $\eta - \phi$ plane. The grid covers $\eta \in [-\pi,\pi]$ and $\phi \in [0,2\pi]$ and has a fine granularity of $0.02\times0.02$. The base features for each pixel are the four-momenta components ($p_x, p_y, p_z, E$) of the particles it contains. If multiple particles fall into the same pixel, their four-momenta are summed. 
To enhance the model's $b$-jet identification capabilities, we include two additional features: the absolute charge $|Q|$ and the transverse impact parameter $d_0$. When multiple particles occupy the same pixel, their $|Q|$ are summed, while the maximum $d_0$ among them is taken. The summation of momenta mimics the energy deposition in a calorimeter, where particles hitting the same cell are indistinguishable. Consequently, pileup particles can contaminate the feature values of jet constituents, making a pileup mitigation strategy essential for accurate momentum reconstruction.
The resulting input for our network is a tensor of shape 
$(315,315,6)$, corresponding to the image height, width, and number of channels.

Our model is trained on a set of ground-truth labels derived for each target ancestor particle. These labels address three distinct tasks:
\begin{itemize}
\item \textbf{Instance Segmentation Mask}: A pixel-wise binary mask that isolates the constituents of a specific jet. For a given ancestor particle, pixels corresponding to its decay products are assigned a value of 1, and all others are set to 0. 
\item \textbf{Object Classification Label}: An integer label that specifies the class of the ancestor particle. The defined classes are: 1 ($t$), 2 ($H$), 3 ($W/Z$), 4 ($b$) and 5 (light-quark/gluon jets). A label of 0 denotes pixels not associated with any targeted jet (i.e., background).
\item \textbf{Kinematic Regression Targets}: A vector of four kinematic variables representing the ancestor particle's four-momentum: transverse momentum ($p_T$), rapidity ($\eta$\footnote{{In order to simplify the representation, $\eta$ will be used to mark both rapidity and pseudo-rapidity in the future. For final-state particles, $\eta$ represents pseudo-rapidity, and for intermediate particles, it represents rapidity.}}), azimuthal angle ($\phi$), and invariant mass ($m$\footnote{The model is trained to regress the true, event-specific invariant mass of a jet rather than a fixed standard value (e.g., the top quark pole mass). This approach is necessary because unstable particles possess a natural decay width, causing their true mass to follow a Breit-Wigner distribution that varies from event to event.}).
\end{itemize}
The ground-truth labels (mask, classification, and kinematics) used for training are generated only for ancestor particles satisfying baseline selections of $p_T>50$ GeV and $|\eta|<\pi$. 

The generation of the segmentation mask for each of these $M$ valid ancestor particles is a multi-step process designed to ensure stability and facilitate model training. 
Firstly, to ensure infrared and collinear (IRC) safety, a pre-selection is applied to the final-state particles assigned to each valid ancestor. A constituent particle is retained only if it meets all the following criteria: transverse momentum $p_T>0.1$ GeV, pseudorapidity $|\eta|<\pi$ and at least 4 other assigned constituents are present within a 20-pixel radius. 
Secondly, for each of the $M$ valid ancestor particles, an initial mask is constructed by aggregating contributions from its set of filtered constituent particles. The procedure for a single constituent particle $i$ at the pixel coordinates $(x^i, y^i)$ is as follows: 1). A 2D Gaussian distribution with a standard deviation of $\sigma=2$ pixels is centered at $(x^i, y^i)$. 2). The Gaussian values are evaluated within a local $9\times9$ pixel patch centered on the particle. 3). These values are then weighted by a factor {$w^i = \ln(p_T^i + e)$} (with $e$ being the base of the natural logarithm to ensure positivity of $w^i$), creating a weighted partial mask for particle $i$. 4). The initial mask for the ancestor is the combination of these partial masks from all its filtered constituents. 
Finally, each of the $M$ initial masks is individually normalized by its own maximum value, ensuring all pixel values lie in the range $[0,1]$. These $M$ normalized masks are then stacked to form the final ground-truth segmentation tensor of shape $(315,315,M)$. 
This elaborate procedure, by creating ``soft", extended target regions through Gaussian smoothing and $p_T$ weighting, can mitigate the challenges associated with training a network on sparse, single-pixel labels, thereby improving learning convergence and stability.

\section{The network architecture} \label{sec3}
\begin{figure}[t!]
\includegraphics[width=0.8\textwidth]{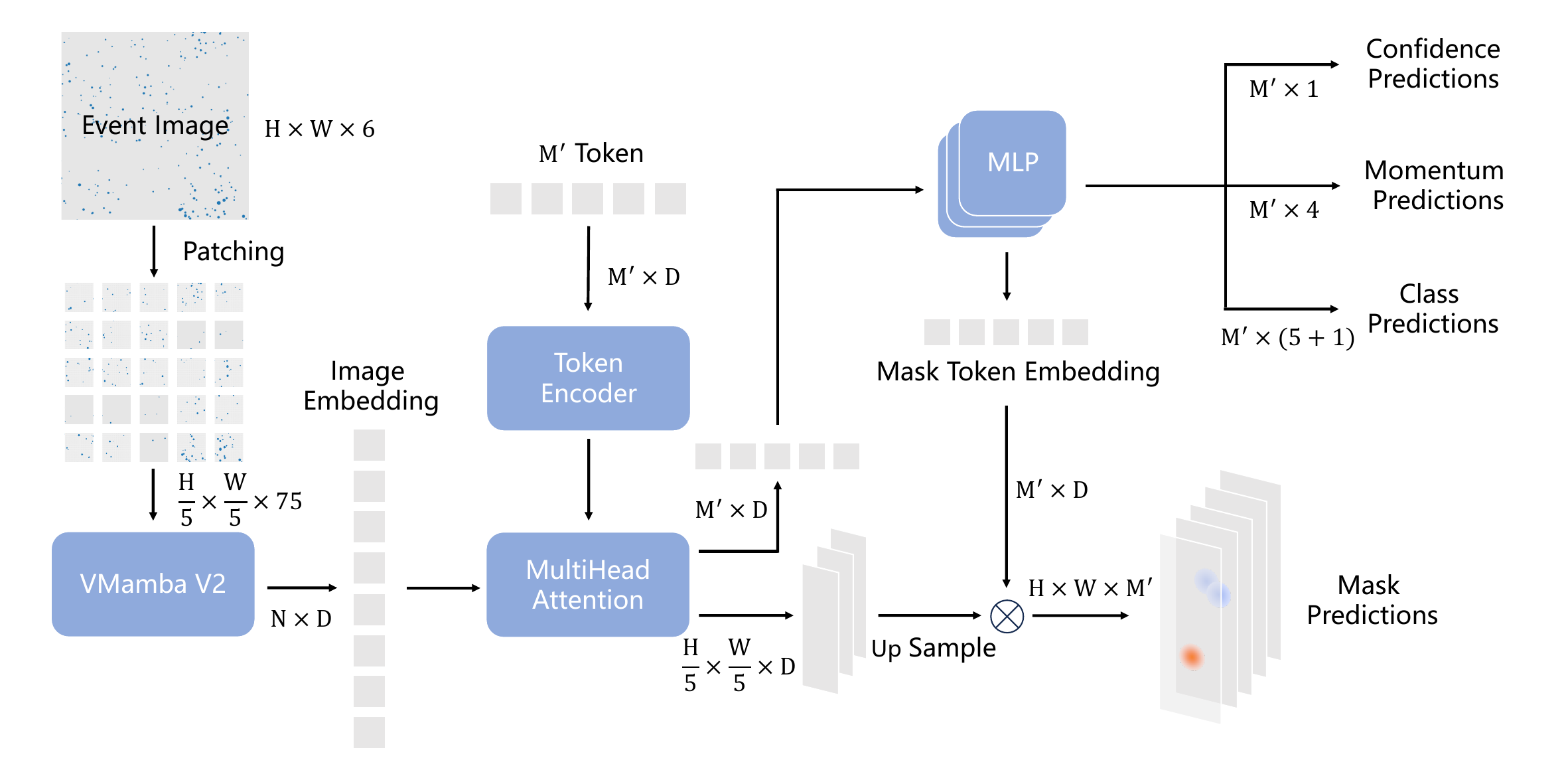}
\caption{Architecture of SAsmJM. The model leverages a VMamba V2 backbone to effectively capture long-range dependencies in jet images. It employs $M^\prime$ learnable Jet Queries Vectors that interact with image features via multi-head attention to decode each instance in parallel. The shared prediction heads directly output the class, kinematics, and confidence for each jet, along with an instance mask generated by a dynamic mask head. This end-to-end design unifies jet segmentation, classification and reconstruction.}
\label{fig:network}
\end{figure}

As illustrated in Figure~\ref{fig:network}, our network employs the VMamba V2 module as its primary feature extractor. Mamba V2 is a state-space model that functions like a recurrent neural network (RNN) but achieves high efficiency through a technique known as ``parallel scanning". This allows for parallelized computation while preserving strong model performance. Initially developed for natural language processing, the Mamba framework has seen rapid adoption in other domains due to its linear-time complexity and powerful performance. VMamba is the variant of this architecture adapted for computer vision. Unlike traditional CNNs, which can struggle to capture long-range dependencies, VMamba excels at establishing connections between distant pixel regions. This capability is particularly crucial for our application, as it allows the model to correctly reconstruct jets that are split across the periodic boundaries of the unrolled detector image.

The input event image is first processed by a ``Patching" module, a stack of convolutional layers that divides the image into a grid of $N = \frac{H}{P} \times \frac{W}{P}$ non-overlapping patches of size $P \times P$, where $H=W=315,\ P = 5$ in our study. The VMamba V2 backbone then processes these patches to produce an image embedding feature with shape $(N, D=256)$. This feature serves as a global embedding for the event, analogous to how a sentence is represented by a vector in natural language processing. This embedding architecture enables a powerful, token-based interaction. For instance, specific query tokens could be used to guide the model's predictions in several ways: 1) providing descriptive context to resolve physical ambiguities, such as the confusion between 2-jet and 3-jet events caused by initial-state radiation; 2) setting kinematic parameters to instruct the model to only consider jets that satisfy certain analysis cuts; or 3) incorporating information from other algorithms, such as anti-$k_T$ jet masks, to prompt a more refined identification.
In the current study, we adopt a simplified token-based approach due to limited computational resources. Our goal is to have the model predict the masks, categories, kinematics, and confidence scores for five distinct jet types. To achieve this, a set of five learnable query tokens, one for each jet type, is introduced. A Multi-Head Attention module fuses the global event embedding with these query tokens, {producing $M^\prime$ contextualized output tokens.} These tokens are then processed by two parallel prediction heads:
\begin{itemize}
\item \textbf{Classification and Kinematics Head}: The output tokens are fed through several MLP layers to produce the final predictions for classification (5 types jet + 1 background), kinematic variables ($p_T, \eta, \phi, m$), and a confidence score for each jet type. 
\item \textbf{Mask Head}: The token embeddings are also used to generate the segmentation masks. Feature maps from the encoder are upsampled via transposed convolutions. The token embeddings are then combined with these upsampled features (e.g., via a dot product) to produce the final mask prediction for each jet.
\end{itemize}

Our model is designed to predict a fixed maximum number of $M^\prime = 32$ potential jets per event image. For a given event, let there be $M (\le M^\prime)$ ground-truth target jets. During training, there is no pre-defined correspondence between the $M^\prime$ predicted jets and the $M$ target jets. To resolve this ambiguity, we formulate the assignment as a bipartite matching problem. We use the Hungarian algorithm to find the optimal one-to-one matching between the set of $M^\prime$ predictions and the set of $M$ ground-truth labels based on a cost function. This procedure yields $M$ matched prediction-label pairs, which are then used to compute the following optimization loss: 

\begin{equation}
\mathcal{L} = \mathcal{L}_{\rm mask} +  \mathcal{L}_{\rm class} + \mathcal{L}_{\rm kin} + \mathcal{L}_{\rm conf}~.~
\end{equation}
The total loss is the sum of four components, each targeting a specific aspect of the prediction task. The segmentation loss is a linear combination of Focal Loss $\mathcal{L}_{\rm FL}$ and Dice Semimetric Loss~\cite{wang2024dicesemimetricloss} $\mathcal{L}_{\rm DML}$: $\mathcal{L}_{\rm mask} = 0.25 \mathcal{L}_{\rm FL} + 0.75 \mathcal{L}_{\rm DML}$. Focal Loss effectively addresses the extreme class imbalance between foreground (jet) and background pixels. Dice Semimetric Loss directly optimizes the overlap, which is well-suited for the soft, Gaussian-smeared target masks used in our training. The classification loss $\mathcal{L}_{\rm class}$ for the jet type is the Multi-Class Focal Loss, this choice is crucial for mitigating the significant class imbalance among the different jet categories, which arise from the 16 different SM processes used for training. By down-weighting well-classified examples, the Focal Loss encourages the model to focus on learning the features of rarer jet types.
The loss for the four-momentum regression is a weighted sum of the L1 loss (\textit{i.e.} the Mean Absolute Error loss) and EIoU loss: $\mathcal{L}_{\rm kin} = 0.8  L1 + 0.2 \mathcal{L}_{\rm EIoU}$. While the L1 loss measures the absolute error, the $\mathcal{L}_{\rm EIoU}$ term provides a scale- and rotation-invariant metric of kinematic agreement. To compute it, we parameterize the kinematics as a pseudo-bounding box: the jet's location $(\eta,\phi)$ is mapped to the box center, and its scale, represented by $p_T$ and $m$, is mapped to the box's width and height. The IoU can be calculated by using the parameters of the predicted bounding box and the truth bounding box. 
The last loss term trains the model's predicted confidence score ($P_{\rm conf}$) to reflect its actual performance on a given prediction. It is defined as
\begin{align}
\mathcal{L}_{\rm conf} = |P_{\rm conf} - \text{score}_{\rm actural}|,
\end{align}
where the target score $\text{score}_{\rm actural}=1/3(P_{\rm class}+\text{IoU}_{\rm mask} + \text{IoU}_{\rm kin})$ is an average of the model's performance on classification, segmentation, and kinematics. The terms $P_{\rm class},~\text{IoU}_{\rm mask}$ and $\text{IoU}_{\rm kin}$ represent the predicted probability for the true class, the mask IoU, and the kinematic pseudo-box IoU, respectively. Crucially, we apply a gradient stop to $\text{score}_{\rm actural}$ so that this loss only updates the weights responsible for producing $P_{\rm conf}$, not the underlying prediction heads. 

Our SAsmJM model, implemented in PyTorch, consists of 4.75 million trainable parameters. The model was trained for 120 epochs on two NVIDIA RTX 3090 (24GB) GPUs, a process that took approximately 194 hours. To ensure robust generalization and prevent overfitting, we used a dedicated validation set to monitor performance throughout training, alongside regularization techniques such as weight decay within the AdamW optimizer. 
We used the AdamW optimizer with a batch size of 40 and a cosine annealing schedule for the learning rate, which started at $2.5 \times 10^{-4}$ and decayed to a final value of $2.5 \times 10^{-6}$. 
Our simulation shows that the validation loss and validation IoU metrics closely tracked the training performance. Both saturated after approximately 100 epochs, indicating that the model does not suffer from significant overfitting.
On a single RTX 3090 GPU, the average inference time per event is approximately 10.6 milliseconds. Each event sample in every epoch was randomly overlaid with pileup interactions, where the number of pileup events was drawn from a Poisson distribution with a mean of $\langle \mu \rangle =50$. To improve the model's rotational invariance and leverage the detector's azimuthal symmetry, we apply a random azimuthal shift during training. This augmentation is performed on-the-fly for each event with a probability of 0.5. 
The shift angle $\Delta\phi = 2\pi z$ is determined by a factor $z$ drawn from a uniform distribution, $z \sim \mathrm{U}(0.1, 0.8)$. 
To maintain consistency, this shift is applied identically to the input image and the ground-truth segmentation masks, while the target kinematic angle $\phi$ is adjusted accordingly.
All shifts respect the $2\pi$ periodicity by wrapping around the image boundaries. A new random shift is generated for each event in every epoch to ensure the model learns from a wide variety of orientations.

To facilitate further research and ensure reproducibility, the code for our neural network and the trained model weights have been made publicly available on GitHub~\cite{sasmjm}.

\section{The network performance} \label{sec4}

The performance of the trained model is evaluated on the test set described in Section II.A. For evaluation, all data augmentations, specifically the random azimuthal shifts, were disabled to ensure reproducible and unbiased performance metrics.

\begin{figure}[htbp]
\includegraphics[width=1.0\textwidth]{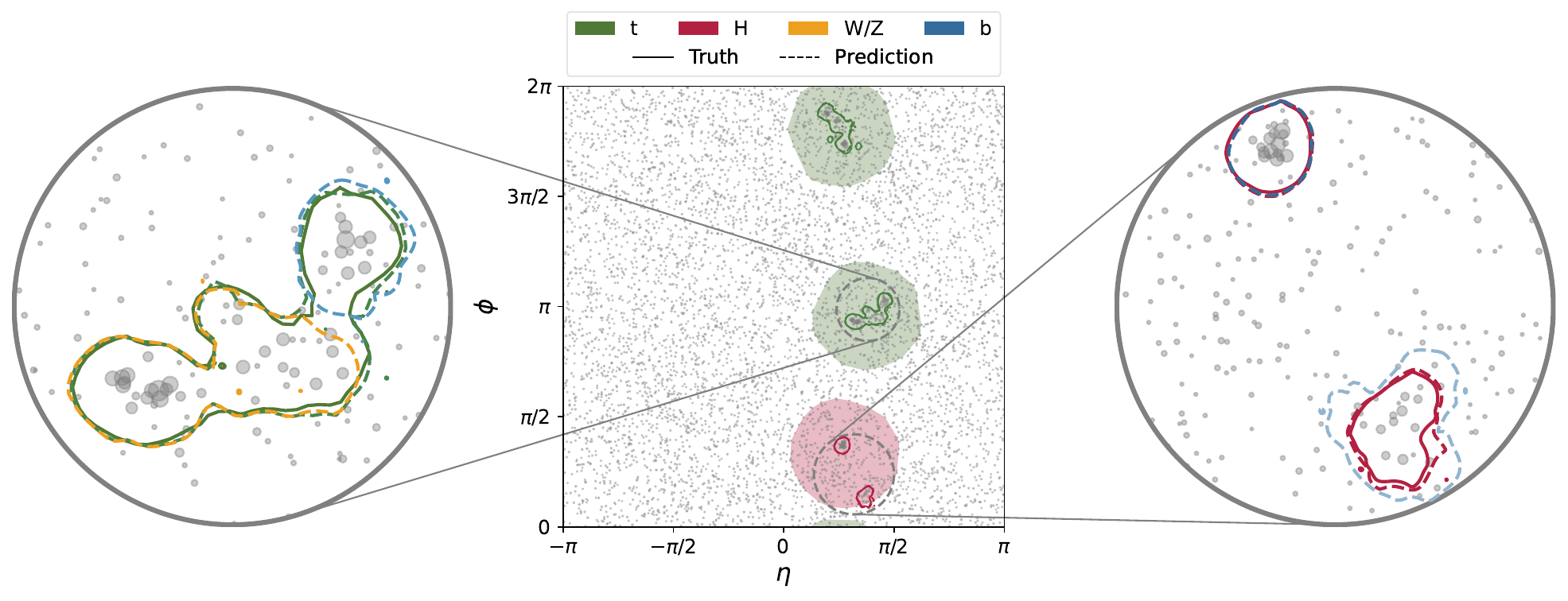}
\caption{Visualization of multi-level jet reconstruction performed by the model on a sample $Ht\bar{t}$ event (with pileup events being superposed on). The image shows all final-state particles (gray circles), with the area of each circle scaled by the particle's logarithmic transverse momentum. Solid and dashed lines represent the boundaries of the ground-truth and predicted masks, respectively. The color of the lines indicates the jet identity (e.g., $H$-jet, top-jet, etc.).
The zoomed-in panels display the sub-jet structure of the primary reconstructed jets. Ground-truth masks for these sub-jets have been omitted from the zoomed-in views for clarity. The shaded regions delineate the jet areas found using the anti-$k_T$ algorithm with cone size parameter $R=0.8$.}
\label{fig:detect}
\end{figure}

To provide an intuitive understanding of the model's performance, Figure~\ref{fig:detect} visualizes its application to a sample $Ht\bar{t}$ event in the $\eta$-$\phi$ plane. 
Final-state particles are represented by gray circles, whose size is proportional to their logarithmic transverse momentum $\mathrm{ln}(p_T)$.  
Our model is designed for multi-level jet identification, which involves simultaneously recognizing not only primary heavy particles (like $H$ boson and top quarks) but also their subsequent decay products (like $W$ bosons and $b$-jets). In the figure, the ground-truth and predicted masks are denoted by solid and dashed lines, respectively, while the colors indicate the jet classification.
The left zoomed-in panel highlights the reconstruction of a top quark decay chain. The model successfully captures the complete $t\to Wb$ decay chain. It identifies the constituent $b$-jet (blue dashed line) and $W$-jet (orange dashed line), and correctly composes them into a parent top-jet (green dashed line). This hierarchical recognition demonstrates a sophisticated understanding of the underlying physics, showcasing a key capability of the simultaneous decomposition of jets into their constituents and the composition of constituents into their parent objects.
The right zoomed-in panel showcases the model's performance in a more challenging topology: a Higgs boson decay where the two resulting $b$-jets are well-separated in the angular plane. Despite the large distance between them, the model excels. It first identifies each energy deposit as an individual $b$-jet (blue dashed lines). Then, demonstrating a grasp of the underlying event kinematics, it correctly associates these non-adjacent jets, grouping them to reconstruct the parent Higgs boson (red dashed line). This capacity to establish long-range correlations and group objects based on physics principles, rather than simple proximity, is a key advantage of our approach and a testament to its sophisticated pattern recognition capabilities. 
For comparison, the figure also displays the jet areas defined by the anti-$k_T$ algorithm with a cone radius of $R=0.8$. In contrast to these larger, fixed-size regions, our model identifies more compact and precise jet boundaries. This precision resolves a key trade-off of fixed-cone algorithms: for the event shown, reducing the anti-$k_T$ cone radius to $R=0.7$ would fail to contain one of the b-jets from the Higgs decay. Our model's ability to define smaller yet complete jet areas is a significant advantage, as it mitigates pileup contamination and thereby enables more accurate momentum regression.

\subsection{The Confusion Matrix}

\begin{figure}[htbp]
\includegraphics[width=0.5\textwidth]{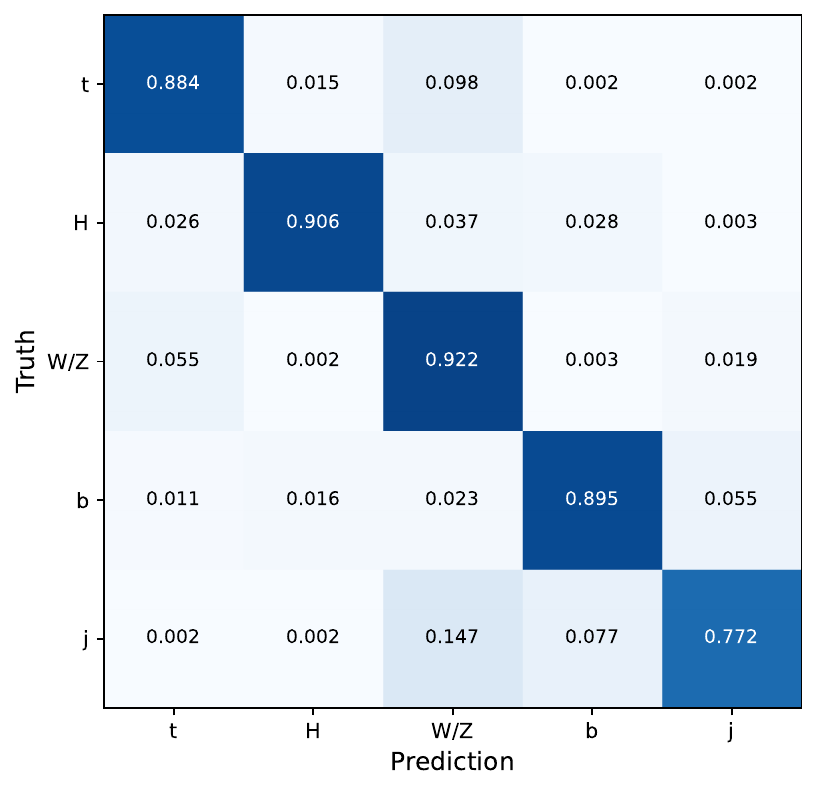}
\caption{Row-normalized confusion matrix summarizing the model's classification performance on five SM jet categories ($t$, $H$, $W/Z$, $b$ and $j$). Row indices represent the true classes, while column indices represent the predicted classes. The diagonal elements represent the recall (true positive rate) for each category, indicating the fraction of jets correctly classified.}
\label{fig:cm}
\end{figure}

Figure~\ref{fig:cm} shows the confusion matrix used to evaluate the model's classification performance. The entries in this matrix are populated using a specific selection and matching procedure. First, we filter the model's predictions, retaining only those with a confidence score greater than 0.3 and a mask IoU greater than 0.5 with at least one ground-truth mask in the event. For each of these surviving predictions, its predicted class is taken as the one with the highest classification score. Its corresponding true class is defined as the class of the ground-truth mask with which it has the highest IoU.

The row-normalized confusion matrix in Figure~\ref{fig:cm} illustrates the model's classification performance. Each row representing a true class sums to one, and the diagonal elements directly correspond to the recall (\textit{i.e.,} true positive rate) for that class. Overall, the model demonstrates robust classification capabilities. High recall rates are achieved for $W/Z$-jets (92.2\%), $H$-jets (90.6\%), $b$-jets (89.5\%), and the top-jet (88.4\%). The lowest recall is for light-jets, at 77.2\%. The most significant confusion occurs where light-jets are misidentified as $W/Z$-jets (14.7\%) or $b$-jets (7.7\%). Additionally, a notable fraction (9.8\%) of top-jets are misclassified as $W/Z$-jets. The remaining off-diagonal elements are small, indicating minimal confusion between other class pairs.

The hierarchy of classification performance directly correlates with the fidelity of the ground-truth labels available for each jet class. The highest recall is achieved for $H$- and $W/Z$-jets, whose constituents can be unambiguously identified from the Monte Carlo truth record without any algorithmic assignment. This provides clean, high-quality training labels. The performance for $t$- and $b$-jets is slightly lower because their ground-truth labels rely on the MILP assignment procedure, which, while powerful, can introduce a small fraction of imperfect labels. The most significant challenge is posed by light-jets, which not only depend on the MILP assignment but also lack a parton-level $p_T$ cut in their generation. This results in a vast and varied kinematic phase space that is inherently more difficult for the network to model, leading to the lowest recall. 
The primary off-diagonal misclassifications also have clear physical interpretations. The tendency for top-jets to be misidentified as $W/Z$-jets is a classic challenge, likely occurring when the model correctly identifies the energetic $W \to q\bar{q}$ decay but fails to associate a softer or merged $b$-jet from the $t \to Wb$ process. More fundamentally, the confusion between light-jets and $W/Z$-jets arises from their potential topological degeneracy. High $p_T$  light-quark or gluon jets often exhibit a two-prong substructure due to radiation that kinematically mimics the signature of a boosted, hadronically decaying $W$ or $Z$ boson, making them intrinsically confusable without additional information.

\subsection{The Momentum Regression}
\begin{figure}[htbp]
\includegraphics[width=0.45\textwidth]{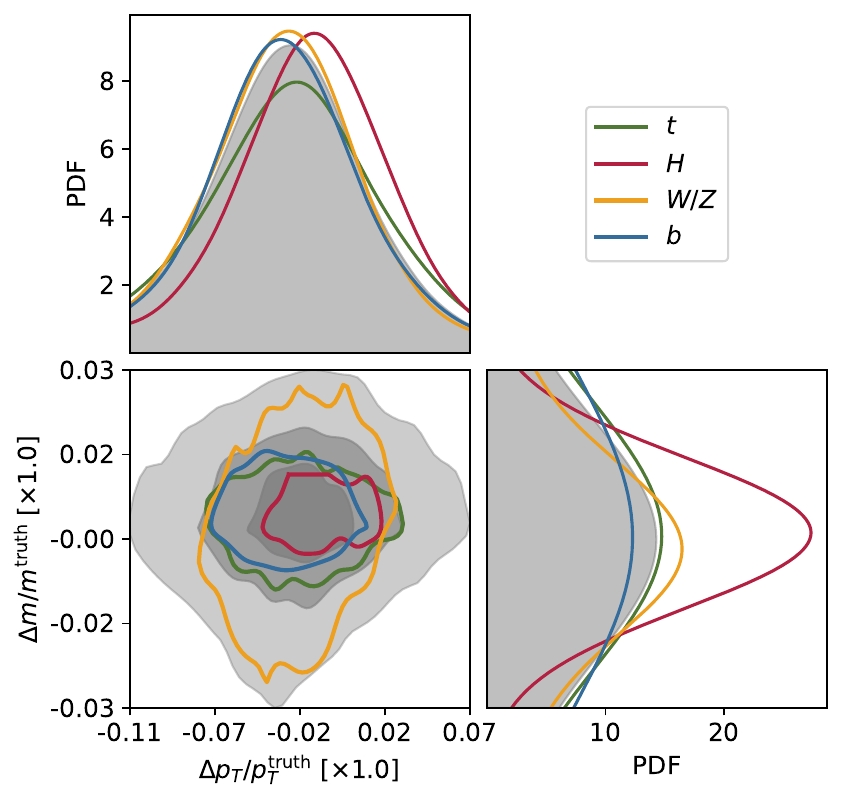}
\includegraphics[width=0.45\textwidth]{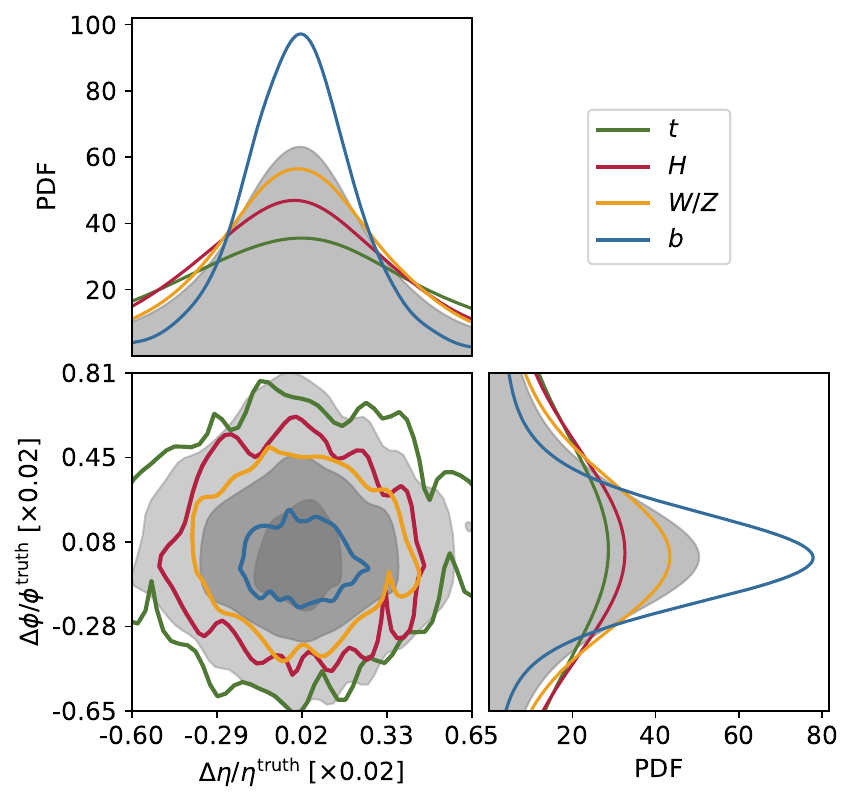}
\caption{Distributions of the kinematic reconstruction error in the $(\Delta p_T/p^{\rm truth}_T, \Delta m/m^{\rm truth})$ plane (left panel) and in the $(\Delta\eta/\eta^{\rm truth}, \Delta\phi/\phi^{\rm truth})$ plane (right panel). The main plot shows the 2-dimensional error distribution. The shaded regions represent contours containing 60\%, 40\%, and 20\% of the total test events, from lightest to darkest. The solid colored lines represent the 40\% containment contours for each individual jet category. The plots on the top and right of each panel show the 1-dimensional marginalized probability distributions each of the variable errors. }
\label{fig:contour}
\end{figure}

We now evaluate the kinematic reconstruction performance using the same matched event sample as for the confusion matrix analysis. Figure~\ref{fig:contour} and Table~\ref{tab:rmsd} present the results, quantifying the model's performance via the fractional errors in four key variables: transverse momentum $\Delta p_T/p^{\rm truth}_T$, invariant mass $\Delta m/m^{\rm truth}$, rapidity $\Delta\eta/\eta^{\rm truth}$, and azimuthal angle $\Delta\phi/\phi^{\rm truth}$. 

\begin{table}[htbp]
\centering
\begin{tabular}{|c|c|c|c|c|c|c|} \hline
& \multirow{2}{*}{20\%} & \multicolumn{4}{c|}{40\%} & \multirow{2}{*}{60\%} \\
\cline{3-6}
& & $t$ & $H$ & $W/Z$ & $b$ & \\
\hline
$(\Delta p_T/p^{\rm truth}_T,\ \Delta m/m^{\rm truth})$ & 0.02914 & 0.0392 & 0.02322 & 0.04411 & 0.03939 & 0.06266 \\
\hline
$(\Delta\eta/\eta^{\rm truth},\ \Delta\phi/\phi^{\rm truth})$ & 0.00401 & 0.01410 & 0.00964 & 0.00819 & 0.00404 & 0.01262 \\ \hline
\end{tabular}
\caption{Performance of the kinematic reconstruction, quantified by the RMSD for the $(\Delta p_T/p^{\rm truth}_T, \Delta m/m^{\rm truth})$ and $(\Delta\eta/\eta^{\rm truth}, \Delta\phi/\phi^{\rm truth})$, respectively. 
For the subset of events within its 40\% containment contour, the RMSD is calculated individually for each jet type.  \label{tab:rmsd}}
\end{table}

Figure~\ref{fig:contour} presents the joint and marginalized distributions of the fractional reconstruction errors, providing a detailed assessment of the model's kinematic precision. The figure displays two corner plots for momentum/mass errors $(\Delta p_T/p^{\rm truth}_T, \Delta m/m^{\rm truth})$ on the left, and angular errors $(\Delta\eta/\eta^{\rm truth}, \Delta\phi/\phi^{\rm truth})$ on the right. In each 2D panel, the filled gray contours enclose the 60\%, 40\%, and 20\% of the full test sample with the smallest reconstruction errors. The compactness and centering of these contours at the origin (0, 0) indicate excellent overall performance. For a class-by-class comparison, the 40\% containment contours for each jet type are overlaid as solid colored lines. The corresponding 1D marginalized error distributions are shown as histograms on the top and right axes of each panel.
For a quantitative summary, Table~\ref{tab:rmsd} lists the Root Mean Square Deviation (RMSD) for each kinematic variable. The columns labeled ``20\%" and ``60\%" show the RMSD calculated for the subset of all test events that fall within the 20\% and 60\% global error contours, respectively. In contrast, the columns labeled ``40\%" provide the RMSD for each individual jet class, calculated using only the events within that class's own 40\% containment contour.

The left panel of Figure~\ref{fig:contour} clearly shows that Higgs jets are reconstructed with the highest precision in the momentum-mass $(\Delta p_T/p^{\rm truth}_T, \Delta m/m^{\rm truth})$ plane. The $H$-jet contour (red line) is the most compact and its marginalized distributions are the sharpest, a finding corroborated by its lowest RMSD value of 0.02322 in Table~\ref{tab:rmsd}.
Conversely, the $W/Z$-jet reconstruction exhibits the poorest resolution in this plane, with the broadest error contour (orange line) and the largest RMSD (0.04411). This is particularly evident in the fractional mass error. We attribute this to the fact that $W$ and $Z$ bosons are treated as a single class in our model. Since the network does not distinguish between them, it naturally struggles to predict a single, accurate mass value, leading to a larger variance in the reconstructed mass.
Turning to the angular reconstruction performance (Figure~\ref{fig:contour}, right panel), $b$-jets are localized with the highest precision. Their error contour $(\Delta\eta/\eta^{\rm truth}, \Delta\phi/\phi^{\rm truth})$ is the most compact (blue line), and the marginalized distributions are the sharpest. This visual evidence is supported by the quantitative results in Table~\ref{tab:rmsd}, where $b$-jets achieve the lowest angular RMSD of 0.00404. This high precision is attributable to the characteristically collimated nature of $b$-jets, which results in a smaller, more localized energy footprint in the detector image. On the other hand, top-jets display the lowest angular resolution, evidenced by their broad error contour (green line) and largest RMSD. This reflects the inherent challenge of localizing a parent particle from a complex, wide-angle decay signature. The multiple, often well-separated, decay products of a top quark create a diffuse jet structure that makes a precise determination of the original top quark's direction more ambiguous.

Overall speaking, the kinematic reconstruction performance is strongly dependent on the jet's physical characteristics. Our analysis highlights two key findings. First, the model demonstrates specialized strengths, achieving the highest precision for momentum and mass reconstruction on Higgs jets, while delivering exceptional angular resolution for $b$-jets. Second, the overall precision is correlated with the quality of the ground-truth labels. The superior performance on $H$-jets, which do not require algorithmic momentum assignment, supports the hypothesis that labeling fidelity is a primary driver of reconstruction accuracy. The degradation in performance from the 20\% to the 60\% global error contours quantifies the error tails and provides a benchmark for the model's robustness across the full event sample.

\subsection{The Precision-Recall Curve}
\begin{figure}[thbp]
\includegraphics[width=0.45\textwidth]{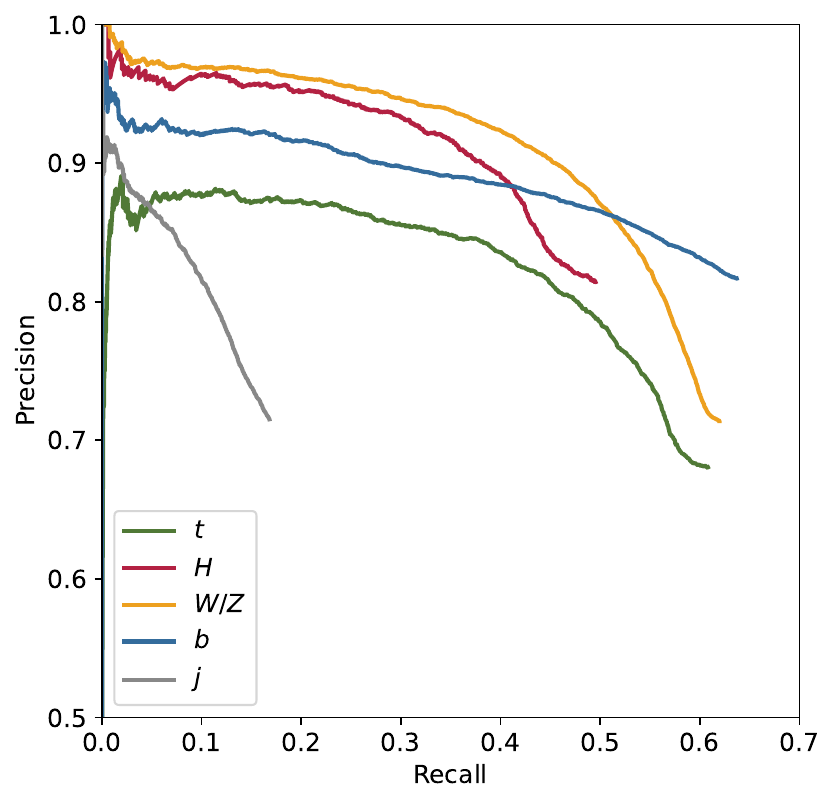}
\caption{Precision-Recall curves for the classification of five SM jet categories: top-jet, Higgs-jet, $W/Z$-jet, $b$-jet, and light-jet. The x-axis represents recall (efficiency), and the y-axis represents precision (purity). The ideal performance is located at the top-right corner (1, 1).\label{fig:pr}}
\end{figure}

To evaluate the classification performance across a range of confidence thresholds, we generated Precision-Recall (PR) curves for each of the five jet classes, as shown in Figure~\ref{fig:pr}. From these curves, we also calculate the Average Precision at an IoU threshold of 0.5 (AP$^{50}$). The procedure to generate these metrics is as follows. First, we consider all model predictions with a confidence score greater than 0.3. For each such prediction, we identify its highest-scoring class. Then, we compute the mask IoU between this prediction's corresponding mask and all ground-truth masks of that same class. The prediction is then definitively matched to the ground-truth mask that yields the highest IoU, forming a single prediction-truth pair for evaluation.
These matched pairs and their associated confidence scores form the basis for the final metrics. The PR curves are then plotted by sweeping a threshold across the confidence scores. The (AP$^{50}$)  is calculated as the area under this curve, with a prediction being classified as a true positive only if its mask IoU with the matched ground-truth object is $\ge 0.5$. This evaluation, which considers performance across all confidence thresholds, provides a more holistic view than the single-operating-point confusion matrix.

\begin{table}[htbp]
\centering
\begin{tabular}{|c|c|c|c|c|c|} \hline
& $t$ & $H$ & $W/Z$ & $b$ & $j$ \\
\hline
AP$^{50}$ & 0.505 & 0.458 & 0.569 & 0.568 & 0.139 \\
\hline
Mask IoU & 0.624 & 0.625 & 0.617 & 0.737 & 0.652 \\ \hline
Kinematic IoU & 0.925 & 0.934 & 0.894 & 0.906 & 0.598 \\ \hline
\end{tabular}
\caption{Quantitative summary of the model's performance on different jet types. The table lists the AP$^{50}$, along with the mean Mask IoU and Kinematic IoU scores, averaged over all successfully matched predictions.
\label{tab:mAP}}
\end{table}

The quantitative metrics in Table~\ref{tab:mAP} confirm the trends observed in the confusion matrix, with $W/Z$, $b$, and top-jets showing high AP$^{50}$ scores, while the light-jet remains the most challenging. However, a more nuanced analysis reveals key trade-offs in the model's learning strategy. For example, despite the $H$-jet's higher peak precision (see Fig.~\ref{fig:pr}), its AP$^{50}$ is lower than the top-jet's. This suggests that while the most confident $H$-jet predictions are purer, the top-jet maintains a more robust performance across all confidence thresholds.

This trade-off between signature ``purity" and reconstruction ``robustness" stems from the jets' intrinsic properties. The Higgs boson's clean $b\bar{b}$ decay allows for high-precision identification of ideal candidates, but its simple topology offers fewer robust features, causing performance to degrade on less distinct examples. In contrast, the top quark's complex $t\to Wb$ decay provides a richer feature set. The model's ability to consistently identify the $W$ and $b$ sub-jets offers a stable, topologically-driven foundation, leading to more robust overall performance. This directly highlights our framework's primary motivation: enhancing reconstruction by learning from rich sub-jet information.

The performance of each jet class is supported by a combination of segmentation and kinematic accuracy. The $W/Z$-jet's high AP$^{50}$ is matched by strong Mask and Kinematic IoU scores. Similarly, the $b$-jet's excellent AP$^{50}$ is underpinned by the highest Mask IoU of any class (0.737), attributable to its highly collimated single-cluster structure. The $H$-jet, despite a slightly lower AP$^{50}$, achieves the best Kinematic IoU (0.934), reinforcing the hypothesis that high-quality, fully-determined ground-truth labels lead to superior kinematic reconstruction.
The top-jet also achieves a remarkably high Kinematic IoU (0.925), indicating that the model effectively learns its distinct mass and the $t\to Wb$ decay topology, overcoming potential noise in the ground-truth labels. However, its overall AP$^{50}$ is limited by its confusion with $W/Z$-jets. Conversely, the light-jet's very low AP$^{50}$ and Kinematic IoU, despite a reasonable Mask IoU, confirms that the model can localize the jet's energy (a low-level task) but struggles with classification and kinematic regression due to the diverse and unconstrained nature of its features (a high-level task).

In summary, our analysis shows that the model consistently excels at the low-level task of segmenting localized energy depositions (high Mask IoU for all classes). However, performance on the higher-level tasks of classification (AP$^{50}$) and kinematic regression (Kinematic IoU) is more sensitive to the complexity of jet signatures and the fidelity of ground-truth labels. Our comprehensive metrics quantitatively illustrate how these factors impact different facets of the model's predictive capabilities.

\section{Test on more general cases} \label{sec5}

We performed two further studies to probe the robustness and generalization power of our model. First, we evaluated its performance stability against varying levels of pileup. Second, we tested its ability to generalize to new physics signals by attempting to reconstruct particles from resonant processes not included in the training data.

\subsection{The pileup effects}

\begin{table}[htbp]
\centering
\begin{tabular}{|c|c|c|c|c|c|c|} \hline
$\langle \mu \rangle$ & 5 & 20 & 50 & 100 & 150 & 200 \\
\hline
$t$ & 0.537 & 0.527 & 0.505 & 0.47 & 0.426 & 0.364 \\ \hline
$H$ & 0.482 & 0.472 & 0.458 & 0.429 & 0.392 & 0.341 \\ \hline
$W/Z$ & 0.604 & 0.591 & 0.569 & 0.519 & 0.466 & 0.401 \\ \hline
$b$ & 0.588 & 0.58 & 0.568 & 0.539 & 0.517 & 0.487 \\ \hline
\end{tabular}
\caption{Robustness of the jet classification performance against pileup. The table give the AP$^{50}$ for each jet class as the mean number of pileup interactions per event $\langle \mu \rangle$ is varied.
\label{tab:pileup}}
\end{table}

As shown in Table~\ref{tab:pileup}, increasing pileup uniformly degrades the classification performance across all jet classes. This underscores the challenge posed by pileup, as the additional energy deposits can distort jet shapes, contaminate substructure observables, and create combinatorial ambiguities. Nevertheless, the degree of performance degradation is not uniform; some jet classes exhibit greater resilience to pileup than others. To quantify this, we define the performance drop as the relative decrease in AP$^{50}$ from the low-pileup $\langle \mu \rangle=5$ to the high-pileup $\langle \mu \rangle=200$ scenario. The $b$-jet exhibits the highest resilience to pileup. Its AP$^{50}$ drops by only 17.2\% (from 0.588 to 0.487). In contrast, the more complex, multi-prong jets show greater sensitivity. The $W/Z$-jet's AP$^{50}$ drops by 33.6\% (from 0.604 to 0.401), and the top-jet shows a similar degradation of 32.2\% (from 0.537 to 0.364). The Higgs jet, also a two-prong decay, is similarly affected, with its AP$^{50}$ decreasing by 29.3\% (from 0.482 to 0.341). 

These varying degrees of performance degradation have clear physical interpretations. The $b$-jet's robustness likely stems from its reliance on features that are inherently resilient to pileup, such as the large transverse impact parameter of its tracks and its simple, collimated energy deposit. These localized features are less affected by the diffuse, low-$p_T$ energy from pileup. In contrast, jets with complex internal structures, like hadronically decaying top- and $W/Z$-jets, are more susceptible. The subtle substructure information crucial for their identification is more easily distorted or obscured by pileup, leading to a more pronounced performance drop. Similarly, the $H$-jet's identification relies on resolving its distinct decay products; pileup contaminates these, making a clean separation more difficult. 
While pileup poses a challenge, we anticipate that the model's high-pileup performance could be further improved by training on a dataset more heavily weighted towards high-pileup scenarios or by incorporating dedicated pileup mitigation techniques.

\subsection{Reconstruction of BSM resonances} 
\begin{figure}[htbp]
\includegraphics[width=0.45\textwidth]{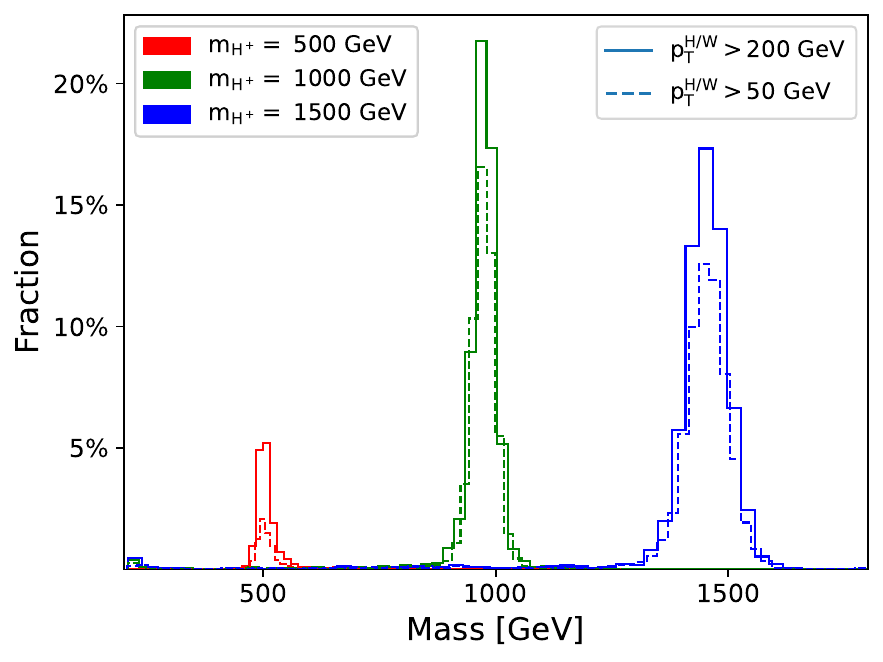}
\includegraphics[width=0.45\textwidth]{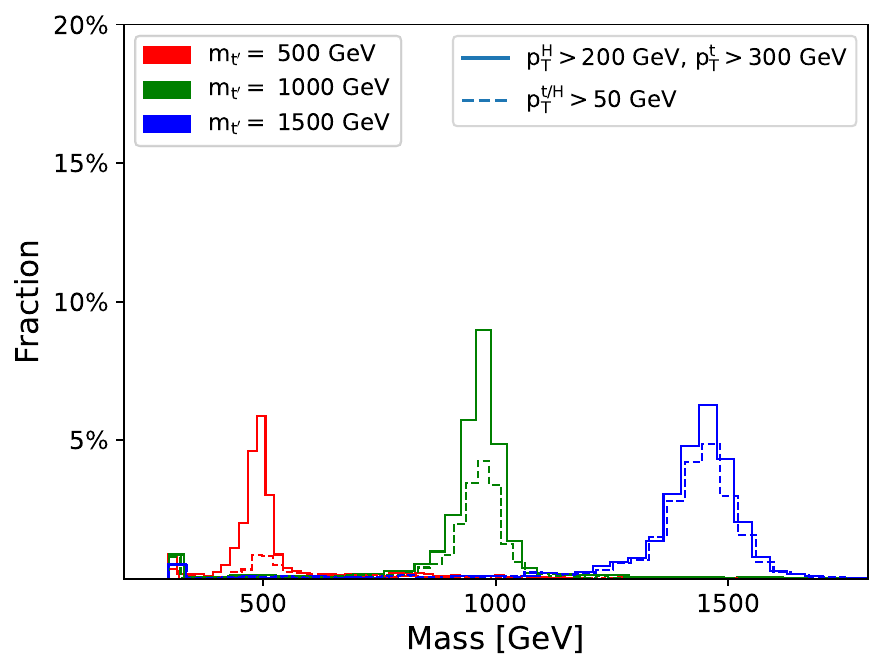}
\caption{Reconstructed invariant mass distributions for two BSM signals: $H^+$ from $pp \to H^+(\to HW)$ (left panel) and $t^\prime$ from $pp \to Zt^\prime(\to tH)$ (right panel). The distributions are shown for three benchmark signal masses: 500, 1000, and 1500 GeV. For each mass point, the solid lines correspond to signal samples generated with the same kinematic cuts as the SM training data, while the dashed lines correspond to samples with relaxed kinematic cuts. All samples are simulated with a mean pileup of $\langle \mu \rangle =50$. 
\label{fig:bsm}}
\end{figure}

To test the model's ability to generalize to new physics, we generated two BSM signal processes for testing: 
$p p \to H^+(\to HW)$ and $p p \to Zt^\prime(\to tH)$ with the $Z$ boson decaying hadronically.
For each process, we created two distinct sets of samples. The first set was generated with the same kinematic cuts as our SM training data. The second set was generated with these cuts relaxed, containing no restrictions at the event generation stage. For each of these four sample sets (two processes $\times$ two cut scenarios), we generated 10000 events and overlaid a mean of $\langle \mu \rangle =50$ pileup events. The reconstruction results for these BSM signals are presented in Figure~\ref{fig:bsm}. 

For both the $H^+$ and $t^\prime$ processes, the reconstructed mass distributions in Figure~\ref{fig:bsm} show clear peaks near the nominal generated mass values (500, 1000, and 1500 GeV). This result is significant, as it demonstrates that the model, despite being trained exclusively on SM processes, can generalize to successfully identify and reconstruct unseen BSM signals. A clear performance difference is observed between samples generated with nominal kinematic cuts (solid lines) and those with relaxed cuts (dashed lines). The nominal-cut samples consistently yield narrower peaks, indicating better mass resolution. This suggests that selecting for high-$p_T$ decay products, which are kinematically similar to the boosted objects in the training data, leads to more accurate mass reconstruction. Conversely, the relaxed-cut samples exhibit broader mass distributions with a prominent low-mass tail. This degradation is expected, as lower-$p_T$ decay products are more susceptible to distortions from pileup and are more likely to be misidentified, thus degrading the overall reconstruction quality.
Furthermore, a clear performance hierarchy is observed between the $H^+$ and $t^\prime$ resonances, {\textit{i.e.}} the $H^+$ resonance is reconstructed with much better resolution than the $t^\prime$. This result directly reflects the performance hierarchy observed in the SM analysis, where the constituent particles ($H$- and $W$-jets) were identified with higher fidelity than the top quark's decay products. This consistency demonstrates that the model's fundamental strengths and weaknesses are preserved when generalizing to new physics.

\section{Conclusion} \label{sec6}
In this work, we have introduced SAsmJM, a novel end-to-end framework for multi-level jet reconstruction in collider events, leveraging the power of the Mamba architecture. By treating entire events as single images and employing the VMamba-V2 backbone, our model effectively captures long-range dependencies with linear complexity, overcoming a key limitation of traditional CNNs. This approach unifies instance segmentation, classification, and kinematic regression into a single multi-task learning system, allowing for the simultaneous identification of both primary jets and their internal sub-jet structures.

Our model demonstrates excellent performance on a diverse set of the SM processes. The classification accuracy is robust, achieving high recall rates for $W/Z$-jets (92.2\%), $H$-jets (90.6\%), $b$-jets (89.5\%), and top-jets (88.4\%). The primary confusions observed are physically motivated, such as the misidentification of light-jets as $W/Z$-jets (14.7\% rate), which arises from their potential topological similarities in the boosted regime. The model exhibits specialized strengths in kinematic reconstruction, achieving the highest precision for the momentum and mass of $H$-jets (RMSD of 0.023 for the best 40\% events) and exceptional angular resolution for $b$-jets (RMSD of 0.004 for the best 40\% events). 

The analysis of the AP$^{50}$ metric provides deeper insights into the model's learning strategy. While $W/Z$-jets (0.569) and $b$-jets (0.568) achieve the highest overall scores, the comparison between $H$-jets (0.458) and top-jets (0.505) is particularly revealing. Despite the $H$-jet's higher peak precision, its performance degrades more rapidly at lower confidence thresholds. In contrast, the top-jet's more complex $t\to Wb$ decay provides a richer, topologically-constrained feature set. The model's ability to learn from the $W$ and $b$ sub-jets leads to a more robust identification across a wider range of scenarios. This finding powerfully validates the core motivation of our multi-level approach: enabling the network to learn the underlying physics of jet decay enhances the overall reconstruction robustness.

Furthermore, our model shows considerable resilience against pileup and strong generalization capabilities. The $b$-jet's performance is the most stable, with its AP$^{50}$ degrading by only 17.2\% when increasing the mean pileup events from 5 to 200. This is attributed to its reliance on localized, pileup-resilient features like its collimated core and associated displaced tracks. 
In contrast, jets defined by their complex substructure ($W/Z$, $H$, $t$) are more sensitive, with performance drops of around 30\%, as pileup contaminates these crucial features. When tested on unseen BSM signals ($H^+$ and $t^\prime$), the model successfully reconstructs their mass peaks, demonstrating that it has learned generalizable physical principles rather than memorizing specific SM processes. The performance hierarchy observed in the SM analysis is preserved in the BSM context, further underscoring this point.

The Mamba-based framework presents a powerful and efficient solution for the complex task of jet reconstruction at the LHC. It successfully performs multi-level identification and demonstrates a sophisticated understanding of jet physics. Future work will focus on further improving pileup mitigation, including more subtle detector effects and so on. The ultimate goal is the application and validation of this framework on real experimental data, where its end-to-end nature and detailed, hierarchical output could provide a rich and comprehensive new tool for discovery and precision measurements at the LHC and future colliders.

\begin{acknowledgments}
The authors gratefully acknowledge the valuable discussions and insights provided by the members of the Collaboration of Precision Testing and New Physics.
This work was supported by the National Natural Science Foundation of China under grant Nos. 11905149, 12335002, and 12275185 and the Natural Science Foundation of Sichuan Province under grant No. 2023NSFSC1329.
\end{acknowledgments}

\bibliographystyle{jhep}
\bibliography{HTdetection}

\providecommand{\href}[2]{#2}\begingroup\raggedright\begin{thebibliography}{10}

\bibitem{Salam:2010nqg}
G.P.~Salam, \emph{{Towards Jetography}},
  \href{https://doi.org/10.1140/epjc/s10052-010-1314-6}{\emph{Eur. Phys. J. C}
  {\bfseries 67} (2010) 637} [\href{https://arxiv.org/abs/0906.1833}{{\ttfamily
  0906.1833}}].

\bibitem{Cacciari:2008gp}
M.~Cacciari, G.P.~Salam and G.~Soyez, \emph{{The anti-$k_t$ jet clustering
  algorithm}}, \href{https://doi.org/10.1088/1126-6708/2008/04/063}{\emph{JHEP}
  {\bfseries 04} (2008) 063} [\href{https://arxiv.org/abs/0802.1189}{{\ttfamily
  0802.1189}}].

\bibitem{Abdesselam:2010pt}
A.~Abdesselam et~al., \emph{{Boosted Objects: A Probe of Beyond the Standard
  Model Physics}},
  \href{https://doi.org/10.1140/epjc/s10052-011-1661-y}{\emph{Eur. Phys. J. C}
  {\bfseries 71} (2011) 1661}
  [\href{https://arxiv.org/abs/1012.5412}{{\ttfamily 1012.5412}}].

\bibitem{Altheimer:2012mn}
A.~Altheimer et~al., \emph{{Jet Substructure at the Tevatron and LHC: New
  results, new tools, new benchmarks}},
  \href{https://doi.org/10.1088/0954-3899/39/6/063001}{\emph{J. Phys. G}
  {\bfseries 39} (2012) 063001}
  [\href{https://arxiv.org/abs/1201.0008}{{\ttfamily 1201.0008}}].

\bibitem{Altheimer:2013yza}
A.~Altheimer et~al., \emph{{Boosted Objects and Jet Substructure at the LHC.
  Report of BOOST2012, held at IFIC Valencia, 23rd-27th of July 2012}},
  \href{https://doi.org/10.1140/epjc/s10052-014-2792-8}{\emph{Eur. Phys. J. C}
  {\bfseries 74} (2014) 2792}
  [\href{https://arxiv.org/abs/1311.2708}{{\ttfamily 1311.2708}}].

\bibitem{Adams:2015hiv}
D.~Adams et~al., \emph{{Towards an Understanding of the Correlations in Jet
  Substructure}},
  \href{https://doi.org/10.1140/epjc/s10052-015-3587-2}{\emph{Eur. Phys. J. C}
  {\bfseries 75} (2015) 409}
  [\href{https://arxiv.org/abs/1504.00679}{{\ttfamily 1504.00679}}].

\bibitem{Larkoski:2017jix}
A.J.~Larkoski, I.~Moult and B.~Nachman, \emph{{Jet Substructure at the Large
  Hadron Collider: A Review of Recent Advances in Theory and Machine
  Learning}}, \href{https://doi.org/10.1016/j.physrep.2019.11.001}{\emph{Phys.
  Rept.} {\bfseries 841} (2020) 1}
  [\href{https://arxiv.org/abs/1709.04464}{{\ttfamily 1709.04464}}].

\bibitem{Kogler:2018hem}
R.~Kogler et~al., \emph{{Jet Substructure at the Large Hadron Collider:
  Experimental Review}},
  \href{https://doi.org/10.1103/RevModPhys.91.045003}{\emph{Rev. Mod. Phys.}
  {\bfseries 91} (2019) 045003}
  [\href{https://arxiv.org/abs/1803.06991}{{\ttfamily 1803.06991}}].

\bibitem{Butterworth:2008iy}
J.M.~Butterworth, A.R.~Davison, M.~Rubin and G.P.~Salam, \emph{{Jet
  substructure as a new Higgs search channel at the LHC}},
  \href{https://doi.org/10.1103/PhysRevLett.100.242001}{\emph{Phys. Rev. Lett.}
  {\bfseries 100} (2008) 242001}
  [\href{https://arxiv.org/abs/0802.2470}{{\ttfamily 0802.2470}}].

\bibitem{Plehn:2010st}
T.~Plehn, M.~Spannowsky, M.~Takeuchi and D.~Zerwas, \emph{{Stop Reconstruction
  with Tagged Tops}},
  \href{https://doi.org/10.1007/JHEP10(2010)078}{\emph{JHEP} {\bfseries 10}
  (2010) 078} [\href{https://arxiv.org/abs/1006.2833}{{\ttfamily 1006.2833}}].

\bibitem{Thaler:2010tr}
J.~Thaler and K.~Van~Tilburg, \emph{{Identifying Boosted Objects with
  N-subjettiness}}, \href{https://doi.org/10.1007/JHEP03(2011)015}{\emph{JHEP}
  {\bfseries 03} (2011) 015} [\href{https://arxiv.org/abs/1011.2268}{{\ttfamily
  1011.2268}}].

\bibitem{Thaler:2011gf}
J.~Thaler and K.~Van~Tilburg, \emph{{Maximizing Boosted Top Identification by
  Minimizing N-subjettiness}},
  \href{https://doi.org/10.1007/JHEP02(2012)093}{\emph{JHEP} {\bfseries 02}
  (2012) 093} [\href{https://arxiv.org/abs/1108.2701}{{\ttfamily 1108.2701}}].

\bibitem{Krohn:2009th}
D.~Krohn, J.~Thaler and L.-T.~Wang, \emph{{Jet Trimming}},
  \href{https://doi.org/10.1007/JHEP02(2010)084}{\emph{JHEP} {\bfseries 02}
  (2010) 084} [\href{https://arxiv.org/abs/0912.1342}{{\ttfamily 0912.1342}}].

\bibitem{Ellis:2009su}
S.D.~Ellis, C.K.~Vermilion and J.R.~Walsh, \emph{{Techniques for improved heavy
  particle searches with jet substructure}},
  \href{https://doi.org/10.1103/PhysRevD.80.051501}{\emph{Phys. Rev. D}
  {\bfseries 80} (2009) 051501}
  [\href{https://arxiv.org/abs/0903.5081}{{\ttfamily 0903.5081}}].

\bibitem{Larkoski:2014wba}
A.J.~Larkoski, S.~Marzani, G.~Soyez and J.~Thaler, \emph{{Soft Drop}},
  \href{https://doi.org/10.1007/JHEP05(2014)146}{\emph{JHEP} {\bfseries 05}
  (2014) 146} [\href{https://arxiv.org/abs/1402.2657}{{\ttfamily 1402.2657}}].

\bibitem{Bertolini:2014bba}
D.~Bertolini, P.~Harris, M.~Low and N.~Tran, \emph{{Pileup Per Particle
  Identification}}, \href{https://doi.org/10.1007/JHEP10(2014)059}{\emph{JHEP}
  {\bfseries 10} (2014) 059} [\href{https://arxiv.org/abs/1407.6013}{{\ttfamily
  1407.6013}}].

\bibitem{Guest:2018yhq}
D.~Guest, K.~Cranmer and D.~Whiteson, \emph{{Deep Learning and its Application
  to LHC Physics}},
  \href{https://doi.org/10.1146/annurev-nucl-101917-021019}{\emph{Ann. Rev.
  Nucl. Part. Sci.} {\bfseries 68} (2018) 161}
  [\href{https://arxiv.org/abs/1806.11484}{{\ttfamily 1806.11484}}].

\bibitem{Albertsson:2018maf}
K.~Albertsson et~al., \emph{{Machine Learning in High Energy Physics Community
  White Paper}},
  \href{https://doi.org/10.1088/1742-6596/1085/2/022008}{\emph{J. Phys. Conf.
  Ser.} {\bfseries 1085} (2018) 022008}
  [\href{https://arxiv.org/abs/1807.02876}{{\ttfamily 1807.02876}}].

\bibitem{Radovic:2018dip}
A.~Radovic, M.~Williams, D.~Rousseau, M.~Kagan, D.~Bonacorsi, A.~Himmel et~al.,
  \emph{{Machine learning at the energy and intensity frontiers of particle
  physics}}, \href{https://doi.org/10.1038/s41586-018-0361-2}{\emph{Nature}
  {\bfseries 560} (2018) 41}.

\bibitem{Larkoski:2024uoc}
A.J.~Larkoski, \emph{{QCD masterclass lectures on jet physics and machine
  learning}}, \href{https://doi.org/10.1140/epjc/s10052-024-13341-0}{\emph{Eur.
  Phys. J. C} {\bfseries 84} (2024) 1117}
  [\href{https://arxiv.org/abs/2407.04897}{{\ttfamily 2407.04897}}].

\bibitem{deOliveira:2015xxd}
L.~de~Oliveira, M.~Kagan, L.~Mackey, B.~Nachman and A.~Schwartzman,
  \emph{{Jet-images \textemdash{} deep learning edition}},
  \href{https://doi.org/10.1007/JHEP07(2016)069}{\emph{JHEP} {\bfseries 07}
  (2016) 069} [\href{https://arxiv.org/abs/1511.05190}{{\ttfamily
  1511.05190}}].

\bibitem{Komiske:2016rsd}
P.T.~Komiske, E.M.~Metodiev and M.D.~Schwartz, \emph{{Deep learning in color:
  towards automated quark/gluon jet discrimination}},
  \href{https://doi.org/10.1007/JHEP01(2017)110}{\emph{JHEP} {\bfseries 01}
  (2017) 110} [\href{https://arxiv.org/abs/1612.01551}{{\ttfamily
  1612.01551}}].

\bibitem{Kasieczka:2017nvn}
G.~Kasieczka, T.~Plehn, M.~Russell and T.~Schell, \emph{{Deep-learning Top
  Taggers or The End of QCD?}},
  \href{https://doi.org/10.1007/JHEP05(2017)006}{\emph{JHEP} {\bfseries 05}
  (2017) 006} [\href{https://arxiv.org/abs/1701.08784}{{\ttfamily
  1701.08784}}].

\bibitem{Macaluso:2018tck}
S.~Macaluso and D.~Shih, \emph{{Pulling Out All the Tops with Computer Vision
  and Deep Learning}},
  \href{https://doi.org/10.1007/JHEP10(2018)121}{\emph{JHEP} {\bfseries 10}
  (2018) 121} [\href{https://arxiv.org/abs/1803.00107}{{\ttfamily
  1803.00107}}].

\bibitem{Moreno:2019bmu}
E.A.~Moreno, O.~Cerri, J.M.~Duarte, H.B.~Newman, T.Q.~Nguyen, A.~Periwal
  et~al., \emph{{JEDI-net: a jet identification algorithm based on interaction
  networks}}, \href{https://doi.org/10.1140/epjc/s10052-020-7608-4}{\emph{Eur.
  Phys. J. C} {\bfseries 80} (2020) 58}
  [\href{https://arxiv.org/abs/1908.05318}{{\ttfamily 1908.05318}}].

\bibitem{Komiske:2018cqr}
P.T.~Komiske, E.M.~Metodiev and J.~Thaler, \emph{{Energy Flow Networks: Deep
  Sets for Particle Jets}},
  \href{https://doi.org/10.1007/JHEP01(2019)121}{\emph{JHEP} {\bfseries 01}
  (2019) 121} [\href{https://arxiv.org/abs/1810.05165}{{\ttfamily
  1810.05165}}].

\bibitem{Qu:2019gqs}
H.~Qu and L.~Gouskos, \emph{{ParticleNet: Jet Tagging via Particle Clouds}},
  \href{https://doi.org/10.1103/PhysRevD.101.056019}{\emph{Phys. Rev. D}
  {\bfseries 101} (2020) 056019}
  [\href{https://arxiv.org/abs/1902.08570}{{\ttfamily 1902.08570}}].

\bibitem{Louppe:2017ipp}
G.~Louppe, K.~Cho, C.~Becot and K.~Cranmer, \emph{{QCD-Aware Recursive Neural
  Networks for Jet Physics}},
  \href{https://doi.org/10.1007/JHEP01(2019)057}{\emph{JHEP} {\bfseries 01}
  (2019) 057} [\href{https://arxiv.org/abs/1702.00748}{{\ttfamily
  1702.00748}}].

\bibitem{Cheng:2017rdo}
T.~Cheng, \emph{{Recursive Neural Networks in Quark/Gluon Tagging}},
  \href{https://doi.org/10.1007/s41781-018-0007-y}{\emph{Comput. Softw. Big
  Sci.} {\bfseries 2} (2018) 3}
  [\href{https://arxiv.org/abs/1711.02633}{{\ttfamily 1711.02633}}].

\bibitem{Andreassen:2018apy}
A.~Andreassen, I.~Feige, C.~Frye and M.D.~Schwartz, \emph{{JUNIPR: a Framework
  for Unsupervised Machine Learning in Particle Physics}},
  \href{https://doi.org/10.1140/epjc/s10052-019-6607-9}{\emph{Eur. Phys. J. C}
  {\bfseries 79} (2019) 102}
  [\href{https://arxiv.org/abs/1804.09720}{{\ttfamily 1804.09720}}].

\bibitem{Semlani:2023kzf}
Y.~Semlani, M.~Relan and K.~Ramesh, \emph{{PCN: a deep learning approach to jet
  tagging utilizing novel graph construction methods and Chebyshev graph
  convolutions}}, \href{https://doi.org/10.1007/JHEP07(2024)247}{\emph{JHEP}
  {\bfseries 07} (2024) 247}
  [\href{https://arxiv.org/abs/2309.08630}{{\ttfamily 2309.08630}}].

\bibitem{Ai:2024mkl}
X.~Ai, W.Y.~Feng, S.-C.~Hsu, K.~Li and C.-T.~Lu, \emph{{Detecting highly
  collimated photon-jets from Higgs boson exotic decays with deep learning}},
  \href{https://arxiv.org/abs/2401.15690}{{\ttfamily 2401.15690}}.

\bibitem{Hammad:2024cae}
A.~Hammad and M.M.~Nojiri, \emph{{Streamlined jet tagging network assisted by
  jet prong structure}},
  \href{https://doi.org/10.1007/JHEP06(2024)176}{\emph{JHEP} {\bfseries 06}
  (2024) 176} [\href{https://arxiv.org/abs/2404.14677}{{\ttfamily
  2404.14677}}].

\bibitem{Mikuni:2025tar}
V.~Mikuni and B.~Nachman, \emph{{Method to simultaneously facilitate all jet
  physics tasks}},
  \href{https://doi.org/10.1103/PhysRevD.111.054015}{\emph{Phys. Rev. D}
  {\bfseries 111} (2025) 054015}
  [\href{https://arxiv.org/abs/2502.14652}{{\ttfamily 2502.14652}}].

\bibitem{Jahin:2025plg}
M.A.~Jahin, S.~Soudeep, A.R.~Aditta, M.F.~Mridha, N.~Fahad and M.J.~Hossen,
  \emph{{Vision Transformers for End-to-End Quark-Gluon Jet Classification from
  Calorimeter Images}},  6, 2025
  [\href{https://arxiv.org/abs/2506.14934}{{\ttfamily 2506.14934}}].

\bibitem{pmlr-v162-qu22b}
H.~Qu, C.~Li and S.~Qian, \emph{Particle transformer for jet tagging},  in
  \emph{Proceedings of the 39th International Conference on Machine Learning},
  K.~Chaudhuri, S.~Jegelka, L.~Song, C.~Szepesvari, G.~Niu and S.~Sabato, eds.,
  vol.~162 of \emph{Proceedings of Machine Learning Research},
  pp.~18281--18292, PMLR, 17--23 Jul, 2022,
  \href{https://proceedings.mlr.press/v162/qu22b.html}{https://proceedings.mlr.press/v162/qu22b.html}
  [\href{https://arxiv.org/abs/2202.03772}{{\ttfamily 2202.03772}}].

\bibitem{Brehmer:2024yqw}
J.~Brehmer, V.~Bres\'o, P.~de~Haan, T.~Plehn, H.~Qu, J.~Spinner et~al.,
  \emph{{A Lorentz-Equivariant Transformer for All of the LHC}},
  \href{https://arxiv.org/abs/2411.00446}{{\ttfamily 2411.00446}}.

\bibitem{Gong:2022lye}
S.~Gong, Q.~Meng, J.~Zhang, H.~Qu, C.~Li, S.~Qian et~al., \emph{{An efficient
  Lorentz equivariant graph neural network for jet tagging}},
  \href{https://doi.org/10.1007/JHEP07(2022)030}{\emph{JHEP} {\bfseries 07}
  (2022) 030} [\href{https://arxiv.org/abs/2201.08187}{{\ttfamily
  2201.08187}}].

\bibitem{Bogatskiy:2022czk}
A.~Bogatskiy, T.~Hoffman, D.W.~Miller and J.T.~Offermann, \emph{{PELICAN:
  Permutation Equivariant and Lorentz Invariant or Covariant Aggregator Network
  for Particle Physics}},  \href{https://arxiv.org/abs/2211.00454}{{\ttfamily
  2211.00454}}.

\bibitem{Ju:2020tbo}
X.~Ju and B.~Nachman, \emph{{Supervised Jet Clustering with Graph Neural
  Networks for Lorentz Boosted Bosons}},
  \href{https://doi.org/10.1103/PhysRevD.102.075014}{\emph{Phys. Rev. D}
  {\bfseries 102} (2020) 075014}
  [\href{https://arxiv.org/abs/2008.06064}{{\ttfamily 2008.06064}}].

\bibitem{Guo:2020vvt}
J.~Guo, J.~Li, T.~Li and R.~Zhang, \emph{{Boosted Higgs boson jet
  reconstruction via a graph neural network}},
  \href{https://doi.org/10.1103/PhysRevD.103.116025}{\emph{Phys. Rev. D}
  {\bfseries 103} (2021) 116025}
  [\href{https://arxiv.org/abs/2010.05464}{{\ttfamily 2010.05464}}].

\bibitem{Li:2020grn}
J.~Li, T.~Li and F.-Z.~Xu, \emph{{Reconstructing boosted Higgs jets from event
  image segmentation}},
  \href{https://doi.org/10.1007/JHEP04(2021)156}{\emph{JHEP} {\bfseries 04}
  (2021) 156} [\href{https://arxiv.org/abs/2008.13529}{{\ttfamily
  2008.13529}}].

\bibitem{Choi:2023slq}
S.K.~Choi, J.~Li, C.~Zhang and R.~Zhang, \emph{{Automatic detection of boosted
  Higgs boson and top quark jets in an event image}},
  \href{https://doi.org/10.1103/PhysRevD.108.116002}{\emph{Phys. Rev. D}
  {\bfseries 108} (2023) 116002}
  [\href{https://arxiv.org/abs/2302.13460}{{\ttfamily 2302.13460}}].

\bibitem{2017arXiv170306870H}
K.~{He}, G.~{Gkioxari}, P.~{Doll{\'a}r} and R.~{Girshick}, \emph{{Mask R-CNN}},
  {\emph{arXiv e-prints} (2017) arXiv:1703.06870}
  [\href{https://arxiv.org/abs/1703.06870}{{\ttfamily 1703.06870}}].

\bibitem{gu2023mamba}
A.~Gu and T.~Dao, \emph{Mamba: Linear-time sequence modeling with selective
  state spaces},  \href{https://arxiv.org/abs/2312.00752}{{\ttfamily
  2312.00752}}.

\bibitem{dosovitskiy2020image}
A.~Dosovitskiy, L.~Beyer, A.~Kolesnikov, D.~Weissenborn, X.~Zhai,
  T.~Unterthiner et~al., \emph{An image is worth 16x16 words: Transformers for
  image recognition at scale}, {\emph{arXiv:2010.11929} (2020) }.

\bibitem{liu2024vmamba}
Y.~Liu, Y.~Tian, Y.~Zhao, H.~Yu, L.~Xie, Y.~Wang et~al., \emph{Vmamba: Visual
  state space model},  \href{https://arxiv.org/abs/2401.10166}{{\ttfamily
  2401.10166}}.

\bibitem{madgraph}
J.~Alwall, R.~Frederix, S.~Frixione, V.~Hirschi, F.~Maltoni, O.~Mattelaer
  et~al., \emph{The automated computation of tree-level and next-to-leading
  order differential cross sections, and their matching to parton shower
  simulations}, \href{https://doi.org/10.1007/jhep07(2014)079}{\emph{Journal of
  High Energy Physics} {\bfseries 2014} (2014) }.

\bibitem{pythia}
T.~Sjöstrand, S.~Mrenna and P.~Skands, \emph{A brief introduction to {PYTHIA}
  8.1}, \href{https://doi.org/10.1016/j.cpc.2008.01.036}{\emph{Computer Physics
  Communications} {\bfseries 178} (2008) 852}.

\bibitem{ATLAS:2012uec}
{ATLAS Collaboration}, \emph{{Summary of ATLAS Pythia 8 tunes}},  Tech. Rep.
  \href{https://cds.cern.ch/record/1474107}{ATL-PHYS-PUB-2012-003}, CERN,
  Geneva (2012).

\bibitem{Skands:2014pea}
P.~Skands, S.~Carrazza and J.~Rojo, \emph{{Tuning PYTHIA 8.1: the Monash 2013
  Tune}}, \href{https://doi.org/10.1140/epjc/s10052-014-3024-y}{\emph{Eur.
  Phys. J. C} {\bfseries 74} (2014) 3024}
  [\href{https://arxiv.org/abs/1404.5630}{{\ttfamily 1404.5630}}].

\bibitem{ATLAS:2016puo}
{ATLAS Collaboration}, \emph{{The Pythia 8 A3 tune description of ATLAS minimum
  bias and inelastic measurements incorporating the Donnachie-Landshoff
  diffractive model}},  Tech. Rep.
  \href{https://cds.cern.ch/record/2206965}{ATL-PHYS-PUB-2016-017}, CERN,
  Geneva (2016).

\bibitem{wang2024dicesemimetricloss}
Z.~Wang, T.~Popordanoska, J.~Bertels, R.~Lemmens and M.B.~Blaschko, \emph{Dice
  semimetric losses: Optimizing the dice score with soft labels},
  \href{https://arxiv.org/abs/2303.16296}{{\ttfamily 2303.16296}}.

\bibitem{sasmjm}
R.~Zhang, \emph{\url{https://github.com/scu-heplab/seg-any-sm-jet}},  2025.

\end{thebibliography}\endgroup
\end{document}